\documentclass[12pt]{article}
\usepackage{amsmath} 
\input{epsf} 
\begin{document} 

\begin{titlepage}
\renewcommand{\thefootnote}{\fnsymbol{footnote}}

\par \vspace{50mm}

\begin{center}
{\Large \bf {Two hadron production in   $e^+ e^-$ annihilation to next-to-leading order accuracy}\footnote{Partially supported by Fundaci\'on Antorchas and CONICET.}}
\end{center}

\par \vspace{8mm}
\begin{center}

{\bf D. de Florian,  L. Vanni }\\

\vspace{5mm}
 Departamento de F\'\i sica, FCEYN, Universidad de Buenos Aires,\\
 (1428) Pabell\'on 1 Ciudad Universitaria, Capital Federal, Argentina

\vspace{5mm}

\end{center}

\par \vspace{2mm}
\begin{center} {\large \bf Abstract} \end{center}
\begin{quote}
\pretolerance 10000

    We discuss the production of two hadrons in  $e^+ e^-$ annihilation within the framework of perturbative $QCD$. The cross section for this process is calculated to next-to-leading order accuracy with a  selection of variables that allows the consideration of events where the two hadrons are detected in the same jet.      In this configuration  we  contemplate  the possibility that the hadrons come  from  a double fragmentation of  a single parton. The double-fragmentation functions 
 required to describe the transition of a parton to two hadrons are also necessary to completely factorize all collinear singularities. We explicitly show that factorization applies to order $\alpha_s$ in the case of two-hadron production.

\end{quote}

\vspace*{\fill}


\end{titlepage}

            
\section{Introduction}

      The  production of one hadron in $e^+e^-$ annihilation has been studied in much detail in perturbative QCD \cite{altarelli B160}. The corresponding cross section for the process $e^+e^- \rightarrow \gamma^*(Q) \rightarrow H(P)+X$ is usually expressed as a function of the variable
\begin{equation}
z =  \frac{2P \cdot Q}{Q^2}
\end{equation}
representing  the energy  fraction carried by the hadron. In this case the cross section can be written as a convolution of the (perturbative computable) partonic cross section $\sigma^i$ and the (non-perturbative) fragmentation functions $D_i^H(x) $ giving the probability of finding a hadron in the parton with 
 momentum fraction $x$, as
\begin{equation}\label{ee H convolucion} 
\frac{d\sigma^H}{dz} =  \sum_i \sigma^i \otimes D_i^H\,. 
\end{equation}

The cross section has been computed to next-to-leading order (NLO) accuracy  in \cite{altarelli B160} and to next-to-next-to-leading order (NNLO) accuracy in \cite{Rijken Neerven}. Furthermore, several analyses of the available data have been performed in the last years and, as a result, fragmentation functions for several hadrons  have been extracted with very good precision.

 Higher order QCD corrections (NLO) to the cross section for the production of  two hadrons in  $e^+e^-$ annihilation have been computed in \cite{altarelli B160} in the particular case when the two hadrons $H_1$ and $H_2$ are selected form different parton jets. 
While a symmetric extension of the one-hadron case to two hadrons would correspond to expressing the differential cross section in terms of the momentum fractions of each hadron defined by
\begin{equation}\label{variables hadronicas}
z_1 =  \frac{2 P_1 \cdot Q}{Q^2}\,\,\,\,\,\,\,\,\,\,\,\,\,\,\,\,\,\,\,\,\,\,\,\,\,\,\,\,\,\,\,\,\,\,\,\,\,\,\,  z_2 = \frac{2 P_2 \cdot Q}{Q^2}\,,
\end{equation}
      the authors in  \cite{altarelli B160} introduced a different set of variables
\begin{equation}\label{varAlt}
z = \frac{2 P_1\cdot q}{Q^2}\,\,\,\,\,\,\,\,\,\,\,\,\,\,\,\,\,\,\,\,\,\,\,\,\,\,\,\,\,\,\,\,\,\,\,\,\,\,\,   u = \frac{P_1 \cdot P_2}{P_1\cdot Q}.
\end{equation}

While $z$ in Eq.(\ref{varAlt}) coincides with $z_1$, the momentum fraction of hadron $H_1$,  the second variable $u$ depends on both the momentum fraction of hadron $H_2$ and the angle $\theta_{12}$ between the hadrons observed from the center of mass system   as
\begin{equation}
u=z_2\,\, \frac{1}{2}(1-\cos \theta_{12})\, ,
\end{equation}
such that  $u$  is approximately  zero when the angle between  the hadrons is small. Therefore, configurations where both hadrons are in the same parton jet  corresponds  to   $u\approx 0$. Consequently, by considering events were $z$ and $u$ are not too small one can ensure that the two hadrons are produced from the hadronization of different partons and the cross section can be reduced to the product of the fragmentation functions $D_i^H$ associated to each hadron \cite{altarelli B160}. In this way, the possibility of a double fragmentation from a single parton is excluded and the expression for the cross section gets simplified.

      In this work we are interested in extending the calculation for the two-hadron cross section  in the full phase space, including the configurations were both hadrons are produced collinearly. In order to be able to consider those events  we will express the cross section in terms of the momentum fractions in Eq.(\ref{variables hadronicas}).

\begin{figure}[htb]
\begin{center}
\begin{tabular}{c}
\epsfxsize=9.0truecm
\epsffile{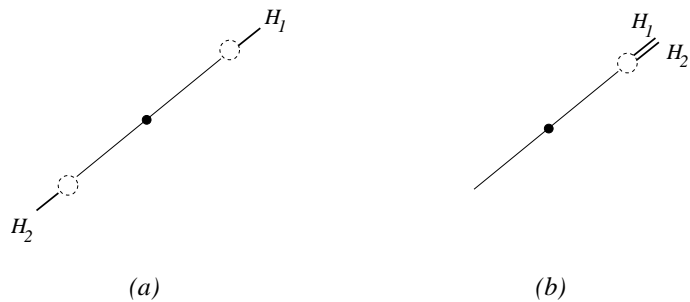}\\
\end{tabular}
\end{center}
\caption{\label{fig11}
{\em  (a) represents hadrons in the first kinematical configuration. (b) represents hadrons in the second configuration, belonging to the same jet.}}
\end{figure}

      With the use of these variables it  is possible to contemplate simultaneously two extreme configurations: the first one  corresponds to the case when the two hadrons are produced in opposite directions (or at least with a clear angular separation) and therefore belonging to different jets (Figure \ref{fig11}a). Hadrons in this configuration can only be originated from the fragmentation of different  partons. The second one corresponds to the case of both hadrons produced  in the same direction,  such  that they are detected in the same jet (Figure \ref{fig11}b).
 In the last case, hadrons could be originated from the fragmentation of  two collinear partons or by the double fragmentation of the single parton. The price one has to pay to fully account for the second configuration is the introduction of a  new set of non-perturbative  phenomenological functions  \cite{mencion de las DDa, mencion de las DD, vendramin 66} to describe the possibility of the transition from a single parton to two hadrons.  

  One needs to introduce, then, {\em the double-fragmentation functions} $DD_p^{H_1H_2}(x_1,x_2)$\footnote{We have slightly modified the original notation introduced in \cite{mencion de las DDa,mencion de las DD} for the double-fragmentation functions to make more noticeable the difference with the usual ones}, as the probability that a parton $p$ fragments into the hadrons $H_1$ and $H_2$ with energy fractions  $x_1$ and $x_2$. The cross section for the production of two hadrons in $e^+e^-$ annihilation can therefore be written  in the following schematic way    
\begin{equation}\label{ee H1H2 convolucion} 
\frac{d\sigma^{H_1 H_2}}{dz_1dz_2} =  \sum_{ij}\sigma^{ij}\otimes D_i^{H_1}\otimes D_j^{H_2} + \sum_{i}\sigma^i \otimes DD_i^{H_1 H_2}, 
\end{equation}
\noindent where $\sigma^{ij}$  is the partonic cross section for the production of partons  $i$ and $j$ and $\sigma^i$ the cross section for  parton $i$.
    The cross section is separated in two terms corresponding to the contribution of the  mechanisms responsible for two-hadron production: single fragmentation of two partons, and double fragmentation of a single parton. 

    At leading order, the first term only contributes to the first configuration, since the two partons that undergo hadronization are produced back-to-back. At next to leading order there is one extra parton which could be emitted collinearly to one of the  others, giving also  origin to hadrons in the second configuration.
Therefore,  at order $\alpha_s$ and beyond, hadrons in the second configuration could be originated from any of the two fragmentation mechanisms, being  not possible to separate the contribution of each term in Eq.(\ref{ee H1H2 convolucion}), unless an additional (unphysical) scale is introduced. Only the sum of both contributions has  physical sense.   
    
    The presence of collinear partons  at order $\alpha_s$  gives origin to collinear  singularities in the cross section, which are manifested in the  form of poles in $\epsilon=(4-N)/2$ when  dimensional regularization is used. By means of the usual redefinition of the the  $D_i^H$ functions, singularities due to collinear partons that give origin to hadrons in the first configuration can be absorbed. However, there appear singularities corresponding to hadrons belonging to the second configuration, originated from collinear partons emitted in the same direction. Since at lowest  order the   $D_i^H$ functions only  participate in processes associated with the first configuration, such singularities can not be absorbed in the single fragmentation term.  We will show that with the  redefinition  of the  $DD_i^{H_1H_2}$ functions in the double fragmentation term all  singularities are factorized. 
 In this sense, the  role played by  the $DD_i^{H_1H_2}$  functions  in $e^+e^-$ annihilation is similar to
 the one of {\em fracture functions} in DIS processes \cite{fractura:c,fractura,fractura:b,fractura:d,Grazzini:1997ih}. For a formal point of view, it is possible to interpret the double-fragmentation functions as the time-like version of fracture functions.
      
  Double-fragmentation functions $DD_i^{H_1H_2}$  fulfill sum rules   in analogy to the sum rules for the usual fragmentation functions  \cite{altarelli B160}. Momentum conservation requires
\begin{equation}\label{relacion sigmaH1H2 y sigmaH1}
\sum_{H_1}\int P_1^{\mu}\, \frac{d\sigma^{H_1H_2}}{dP_1 dP_2}\, dP_1=(Q^{\mu}-P_2^{\mu}) \frac{d\sigma^{H_2}}{dP_2}\, ,
\end{equation}
\noindent being $Q$ the initial total momentum,
 where the right hand side is proportional to  the total free momentum available  for the production of  hadron $H_2$. In particular, energy conservation implies  \cite{mencion de las DD,vendramin 66}
\begin{equation}
\sum_{H_1}\int_0^{1-z_2} dz_1\, z_1\,  DD_i^{H_1H_2}(z_1,z_2) = (1-z_2) D_i^{H_2}(z_2),
\end{equation}
relating the second moment of the double-fragmentation function to the single one.


\section{Two-hadron production in $e^+e^-$}

   In order to formalize the convolution products in Eq.(\ref{ee H1H2 convolucion}) we define the partonic energy fractions associated to each fragmentation mechanism. For the single fragmentation term  two partons  fragment independently with a fraction of the parent parton energy given by  
 \begin{equation}
x_i =  \frac{2p_i \cdot Q}{Q^2},
\end{equation}
 with $p_i$ the momentum of the  parton $i=1,2$. At leading order both variables are fixed to one since no extra gluon radiation is allowed. 

 The  convolution product in the single fragmentation term  of Eq.(\ref{ee H1H2 convolucion}) is expressed in terms of a double integral in $x_1$ and $x_2$ with  integration intervals determined by the kinematical region  allowed  for the partonic process.
This implies
  \begin{eqnarray}\label{region cinematica convolucion}
 0\le & x_1 & \le 1  \,\,\,\,\,\,\,\,\,\,{\rm and} \,\,\,\,\,\,\,\,\,\,\,\, z_1\leq x_1\nonumber \\
1-x_1\le & x_2 & \le 1  \,\,\,\,\,\,\,\,\,\,{\rm and} \,\,\,\,\,\,\,\,\,\,\,\, z_2\leq x_2\, .
\end{eqnarray}

\begin{figure}[htb]
\begin{center}
\begin{tabular}{c}
\epsfxsize=7.5truecm
\epsffile{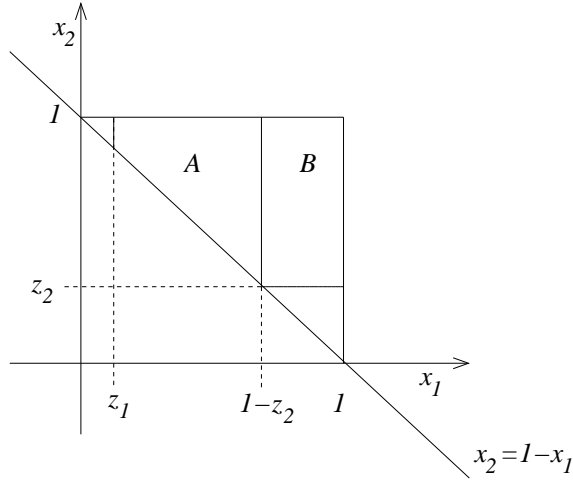}\\
\end{tabular}
\end{center}
\caption{\label{fig14}
{\em Integration regions for the variables $x_1$ and $x_2$ in the single fragmentation term.}}
\end{figure}

The integration zone for the single fragmentation term has to be divided into the two regions $A$ and $B$ indicated in Fig.\ref{fig14}.  

In the case of the double fragmentation term only  one  parton fragments. We define the partonic variable as usual by  
\begin{equation}
x =  \frac{2p \cdot Q}{Q^2},
\end{equation}
with $p$ being the momenta of the fragmenting  parton. With this,  it is possible to write the second term of Eq.(\ref{ee H1H2 convolucion}) as a single convolution  product  with integration limits coming from the request $z_1/x+z_2/x \le 1$.

Using those conditions we can express Eq.(\ref{ee H1H2 convolucion}) as
  \begin{eqnarray}\label{sigmaDxD general}
\frac{d\sigma^{H_1 H_2}}{dz_1dz_2} & = & \sum_{ij} \int_{z_1}^{1-z_2}\frac{dx_1}{x_1} \int_{1-x_1}^1  \frac{dx_2}{x_2}{\frac{d\sigma^{ij}}{dx_1dx_2}}_A D_i^{H_1} \Big (\frac{z_1}{x_1} \Big ) D_j^{H_2} \Big (\frac{z_2}{x_2} \Big )\nonumber \\ \nonumber\\ & + & \sum_{ij}\int_{1-z_2}^1 \frac{dx_1}{x_1} \int_{z_2}^1 \frac{dx_2}{x_2}{\frac{d\sigma^{ij}}{dx_1dx_2}}_BD_i^{H_1} \Big (\frac{z_1}{x_1} \Big ) D_j^{H_2} \Big (\frac{z_2}{x_2} \Big ) \nonumber \\ \nonumber \\ &+ & \sum_{i} \int_{z_1+z_2}^1 \frac{dx}{x^2}\, \frac{d\sigma^i}{dx} DD_i^{H_1 H_2}\Big (\frac{z_1}{x},\frac{z_2}{x} \Big )\, ,
\end{eqnarray}
where to NLO accuracy
\begin{eqnarray}
\frac{d\sigma^i}{dx} &=& {\frac{d\sigma^i}{dx}}^{(0)} + \frac{\alpha_s}{2\pi}\,\, {\frac{d\sigma^i}{dx}}^{(1)}\nonumber \\ \nonumber\\
{\frac{d\sigma^{ij}}{dx_1dx_2}}_K & =&  {\frac{{d\sigma^{ij}}^{(0)}}{dx_1dx_2}}_K+ \frac{\alpha_s}{2\pi}\,\, {\frac{{d\sigma^{ij}}^{(1)}}{dx_1dx_2}}_K,
\end{eqnarray}
 and  $K=A,B$ indicating the integration zone in the single fragmentation term. In Eq.(\ref{sigmaDxD general}) we have  considered the case when $1-z_2\ge z_1$. This implies $z_1+z_2\le 1$, which  corresponds to the kinematical region where the double fragmentation mechanism can also contribute.
   If  $z_1+z_2>1$ the  cross section is reduced only to the second term of 
Eq.(\ref{sigmaDxD general}).   

Some of the partonic cross sections obey symmetry relations that allow to reduce the number of independent quantities to be computed. 
Due to invariance under charge conjugation
\begin{equation}
\frac{d\sigma^{i q}}{dx_1dx_2}=\frac{d\sigma^{i \bar q}}{dx_1dx_2}\,\,\,\,\,\,\,\,\,\,\,\,\,\,\,\,\,\,\,\,\,\,\,\,\,\,
\frac{d\sigma^{q i}}{dx_1dx_2}=\frac{d\sigma^{\bar q i}}{dx_1dx_2}\, .
\end{equation}

To NLO accuracy it is necessary to obtain  only three different partonic cross section $d\sigma^{q \bar q}/{dx_1dx_2}$, $d\sigma^{q g}/{dx_1dx_2}$ and $d\sigma^{g q}/{dx_1dx_2}$.

At leading order the only non-vanishing terms  are \footnote{In this work we restrict the analysis to the case of pure $\gamma^*$ exchange. The extension to $Z$-boson production can be easily performed}
\begin{eqnarray}
{\frac{d\sigma^q}{dx}}^{(0)}  & =& e_q^2\, \sigma_0  \delta(1-x), \nonumber \\ \nonumber \\ {{\frac{{d\sigma^{q \bar q}}^{(0)}    }{dx_1dx_2}}_K} & =& e_q^2\, \sigma_0 \delta(1-x_1)\,\delta(1-x_2). 
\end{eqnarray}
 where $\sigma_0  =\frac{ 4 \pi \alpha_s^2}{3 Q^2}$.

\begin{figure}[htb]
\begin{center}
\begin{tabular}{c}
\epsfxsize=15.0truecm
\epsffile{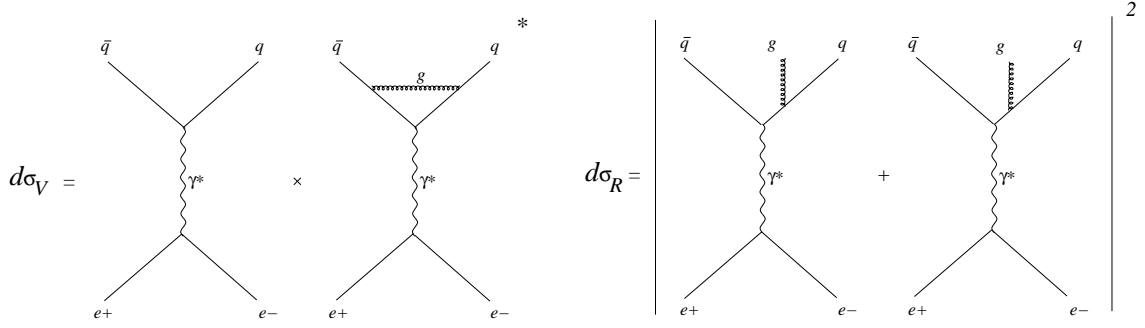}\\
\end{tabular}
\end{center}
\caption{\label{dVdR}
{\em  Virtual and real diagrams contributing to order $\alpha_s$.}}
\end{figure}

 The partonic cross section at order  $\alpha_s$  is  obtained by evaluating the real and virtual  diagrams indicated in Fig.\ref{dVdR} and integrating over the phase space of the final partons expressed in terms of  $x_1,x_2$, such that $d\sigma^{ij}= d\sigma_R + d\sigma_V$.
  We compute the metric and longitudinal contributions to the partonic cross  section obtained, as usual, by replacing the sum over the polarization states of the virtual photon by the  corresponding projectors 
\begin{eqnarray}
P^{(M)}_{\mu\nu} & = & - g_{\mu \nu},\label{proyector metrico}\\ \nonumber \\
P^{(L)}_{\mu\nu} & = & \frac {Q^2}{(p_2\cdot Q)^2} \,p_{2\,\mu}\,p_{2\, \nu}.\label{proyector longitudinal}
\end{eqnarray}

The longitudinal contribution has been calculated projecting in the direction of hadron $H_2$.  In the following, we will present in detail the results  for the metric contribution, since  at NLO singularities of interest occur only on that projection of the cross section. Using dimensional regularization \cite{regularizacion,regularizacion:a}  we obtain for the real part 
\begin{eqnarray}\label{dsigma R  qq H1H2}
d\sigma_R^{(M)}  &=&  e_q^2\, \sigma_0 \frac{\alpha_s}{2\pi} C_F \left (\frac{4\pi \mu^2}{Q^2} \right )^{\epsilon}\frac{1}{\Gamma(2 - \epsilon)}   \left (\frac{1-z^2}{4}\right )^{-\epsilon} x_1^{-2\epsilon}\, x_2^{-2\epsilon}  F(x_1,x_2) dx_1 dx_2,
\end{eqnarray}
with
\begin{equation}\label{Fx1x2}
F(x_1,x_2) = \left [\left (1 -\epsilon \right )^2 \frac{x_1^2 + x_2^2}{(1-x_1)(1-x_2)} -2\epsilon \left (1 - \epsilon \right) \frac{(2-2x_1 - 2x_2 + x_1 x_2)}{(1-x_1)(1-x_2)} \right] \, , 
\end{equation}
and $\mu$ being the dimensional regularization scale. The result for the virtual contribution is
\begin{eqnarray}\label{dsigma V qq H1H2}
d\sigma_V  &=&  e_q^2\, \sigma_0 \frac{\alpha_s}{2\pi} C_F
 \left (\frac{4\pi \mu^2}{Q^2} \right )^{\epsilon}  \frac{\Gamma(1+\epsilon)\Gamma^2 (1-\epsilon)}{\Gamma(1-2\epsilon)}\left [
-\frac{2}{\epsilon^2}  - \frac{3}{\epsilon} -8 + \pi^2   \right]  \delta(1-x_1) \delta(1-x_2)dx_1 dx_2.\,\,\,\,\,\,\,\,\,\,
\end{eqnarray}

  In the previous equations we have labeled the quark as parton $1$ and the anti-quark as parton $2$. The virtual part does not exhibit singularities beyond those already regularized in the form of poles in $\epsilon$. 
In the real part the  divergences appear when the denominators of the function $F(x_1,x_2)$ vanish. These infrared divergences  can be regularized by means of the usual $+$ prescription, which can be easily implement by multiplying and dividing by $(1-x_i)^{1+\epsilon}$ and considering the distribution 
\begin{equation}\label{distribucion e}
(1-x_i)^{-1-\epsilon}\,=\,\,-\frac{1}{\epsilon}\delta(1-x_i) + {\frac{1}{(1-x_i)}_{+[0,\, \underline{1}]}} - \epsilon \left (\frac{\log(1-x_i)}{1-x_i} \right)_{+[0,\, \underline{1}]} + {\cal O}(\epsilon^2) \, ,
\end{equation}
where $F(x)_{+[a, \,\underline{b}]}$ is defined as usual by
\begin{eqnarray}
\int_a^b dx\, f(x)\, F(x)_{+[a, \,\underline{b}]} & = & \int_a^b dx\, [f(x)- f(b)]\,  F(x).\nonumber
\end{eqnarray}

The  range of integration is  indicated as a subscript; furthermore, the subtraction point is underlined.


   For ${d\sigma^{q\bar q}}^{(1)}/dx_1dx_2$ the  singularities of the function $F(x_1,x_2)$ occur  at  $x_2=1$ in zone $A$, and at $x_1=1$ and $x_2=1$ in zone $B$. Applying the  $+$  prescription  as  indicated, the following expression is reached  
 \begin{eqnarray}\label{sigma qq A}
{\frac{{d\sigma^{q\bar q}}^{(1)}}{dx_1dx_2}_K}^{(M)} & = & \sigma_0 e_q^2 \Bigg [ P_{qq}(x_1) \log \left(\frac{Q^2}{\mu^2}\right )\delta(1-x_2)+ P_{qq}(x_2) \log \left(\frac{Q^2}{\mu^2}\right )\delta(1-x_1) \nonumber \\ \nonumber\\ & + &  \frac{1}{\hat{\epsilon}}P_{qq}(x_1) \delta(1-x_2)+ \frac{1}{\hat{\epsilon}} P_{qq}(x_2) \delta(1-x_1)+  f_{qq K}^{(M)}(x_1,x_2) \Bigg ]\, ,
\end{eqnarray}
where $1/\hat{\epsilon}= - 1/\epsilon (4\pi)^{\epsilon} \Gamma(1-\epsilon)/\Gamma(1-2\epsilon) = (-1/\epsilon+\gamma_E-\log 4\pi) +{\cal O}(\epsilon)$,
$P_{ij}(x)$ are the usual Altarelli-Parisi splitting kernels \cite{altarelli B126}
 and the functions ${f_{ij}^M}_K$  are presented in the Appendix.
 

   The $qg$ partonic cross section ${d\sigma^{qg}}^{(1)}/dx_1dx_2$ can be obtained from the $q\bar q$ one relabeling the parton indexes by  a $x_2\rightarrow 2-x_1-x_2$ substitution in the matrix elements. In this case  $F(x_1,x_2)\rightarrow F(x_1, 2-x_1-x_2)$ which develops singularities at  $x_1=1$ in  zone $A$,  and at  $x_1+x_2=1$ in zone $B$. Proceeding like in the previous case and ignoring terms containing distributions without support in the analyzed zones, we obtain  
   \begin{eqnarray}\label{sigma qg A}
{\frac{{d\sigma^{qg}}^{(1)}}{dx_1dx_2}_K}^{(M)}  &= & \sigma_0 e_q^2 \Bigg [ \hat{P}_{qq}^g(x_1) \log \left(\frac{Q^2}{\mu^2}\right )\delta(x_1+x_2-1) + \hat{P}_{gq}^q(x_2) \log \left(\frac{Q^2}{\mu^2}\right )\delta(1-x_1)+ 
\nonumber \\ \nonumber \\ 
&+&  \frac{1}{\hat{\epsilon}} \hat{P}_{gq}^q(x_1)\delta(x_1+x_2-1) + \frac{1}{\hat{\epsilon}} \hat{P}_{gq}^q(x_2)\delta(1-x_1)+ f_{qg K}^{(M)}(x_1,x_2) \Bigg ] \, ,
\end{eqnarray}
where  the functions $\hat{P}_{ji}^k$ are the real LO Altarelli-Parisi kernels, with the index $k$ labeling the third particle in the vertex $i\rightarrow jk$.


   In the $gq$ partonic cross section ${d\sigma^{gq}}^{(1)}/dx_1dx_2$,  $x_1$ is assigned to the gluon and $x_2$ to the quark. Performing the substitution $x_2\leftrightarrow x_1$ in the matrix element used in ${d\sigma^{qg}}^{(1)}/dx_1dx_2$,  $F(x_1, 2-x_1-x_2)\rightarrow F(x_2, 2-x_1-x_2)$ which develops  singularities at $x_2=1$ and $x_1+x_2=1$ in zone $A$, and at $x_2=1$ in zone $B$. The singularities in  zone $A$ will give origin to two different distributions, 
one  associated to the singularity at $x_2=1$ and another to the singularity at $x_1=1-x_2$. The result can be expressed as
\begin{eqnarray}\label{sigma gq A}
 {\frac{{d\sigma^{gq}}^{(1)}}{dx_1dx_2}_K}^{(M)} & = & e_q^2\, \sigma_0  \Bigg [\hat{P}_{gq}^q(x_1) \log \left(\frac{Q^2}{\mu^2}\right )\delta(1-x_2) +   \hat{P}_{qq}^g(x_2) \log \left(\frac{Q^2}{\mu^2}\right ) \delta(x_1+x_2-1)
\nonumber \\ \nonumber \\
 & + & \frac{1}{\hat{\epsilon}} \hat{P}_{gq}^q(x_1)\delta(1-x_2) + \frac{1}{\hat{\epsilon}} \hat{P}_{gq}^q(x_2)\delta(x_1+x_2-1)+  f_{gq K}^{(M)}(x_1,x_2) \Bigg ].
\end{eqnarray}

 This result completes the presentation of the partonic cross sections  that participate in the single fragmentation term.
 
For the double fragmentation term, the cross sections for the production of  a single parton $d\sigma^i/dx$ are required. These are exactly the same as the ones appearing in one-hadron production \cite{altarelli B160}.
 As a  cross-check of our calculation we have re-obtained those coefficients by applying the momentum conservation relation in Eq.(\ref{relacion sigmaH1H2 y sigmaH1}) resulting in
\begin{eqnarray}
 {\frac{d\sigma^{q (1)}}{dx}}^{ (M)}  &=& {\frac{d\sigma^{\bar q (1)}}{dx}}^{ (M)}   =e_q^2\, \sigma_0 \bigg [P_{qq}(x)\log \left (\frac{Q^2}{\mu^2}\right )+ \frac{1}{\hat{\epsilon}} P_{qq}(x) + f^{(M)}_q(x) \bigg ], \nonumber \\ \nonumber \\ 
{\frac{d\sigma^{g (1)}}{dx}}^{ (M)}  &=&  2 e_q^2\,\sigma_0  \bigg [ P_{gq}(x)\log \left (\frac{Q^2}{\mu^2}\right ) + \frac{1}{\hat{\epsilon}} P_{gq}(x) +  f^{(M)}_g(x) \bigg ],
\end{eqnarray}
with $f^{(M)}_i$ given in the Appendix.  

As indicated above, the longitudinal part does not contribute to the singular structure of the cross section to NLO accuracy. The corresponding NLO corrections to the single $f^{(L)}_{ij}$ and double $f^{(L)}_i$ fragmentation  mechanism are shown in the Appendix.


\section{Factorized fragmentation functions}

     The factorization of the bare  fragmentation functions $D_i^H$ at NLO is done in the $\overline{MS}$ factorization scheme in the standard way \cite{altarelli B160}. The expression for the bare  functions in terms of the factorized  ones at the scale $M^2$  is
\begin{eqnarray}\label{ecuaciones de renormalizacion D}
D_i^H(z)= \int_z^1 \frac{du}{u}  \Bigg [\delta(1-u) \delta_{ij} +       \frac{\alpha_s}{2\pi} \bigg [ \log \left (\frac{\mu^2}{M^2} \right )- \frac{1}{\hat{\epsilon}}  \bigg ]P_{ji}(u) \Bigg ] D_j^{H(NLO)}\Big (\frac{z}{u},M^2 \Big)\, ,
\end{eqnarray}
where the factorized distributions are labeled by the upper index (NLO).

It is easy to notice that  not all the singularities in the partonic cross-section are canceled after the factorization of the fragmentation functions is
performed. Singularities belonging to ${d\sigma^{qg}}^{(1)}/dx_1dx_2$ and ${d\sigma^{gq}}^{(1)}/dx_1dx_2$ in zone $A$ still  remain. They correspond to terms with $1/\epsilon$ poles  proportional  to  $\delta(x_1+x_2-1)$, arising from the hadronization of a gluon being emitted collinear to a quark (or anti-quark) that also undergoes hadronization,  giving as a final product two hadrons in the same  jet.  Those singularities clearly cannot be absorbed by the factorization of the $D_i^H$ functions,  since a configuration with two collinear hadrons is not allowed in the single fragmentation term at the lowest order. However, this is exactly the configuration corresponding to the  double fragmentation term, indicating that these singularities could be absorbed by the appropriate factorization of the $DD_i^{H_1H_2}$  functions.  

The expression for the bare double-fragmentation functions $DD_i^{H_1H_2 }(x,y)$ in terms of the NLO factorized ones $DD_i^{H_1H_2(NLO)}(x,y,M^2)$ can be obtained by requiring that all remaining collinear singularities in the partonic cross section are absorbed into the factorized distributions. The expression in the  $\overline{MS}$ factorization scheme, valid to ${\cal O}(\alpha_s)$, is
\begin{eqnarray}
\label{ecuaciones de renormalizacion DD}
DD_i^{H_1H_2}(x,y) &=& \int_{x+y}^1 \frac{du}{u^2} \Bigg [\delta(1-u)\delta_{ij} +\frac{\alpha_s}{2\pi} \bigg [\log \left (\frac{\mu^2}{M^2} \right ) - \frac{1}{\hat{\epsilon}}   \bigg ]P_{ji}(u) \Bigg ] DD_j^{H_1H_2 (NLO)}\Big (\frac{x}{u},\frac{y}{u}, M^2 \Big )\nonumber\\ \nonumber \\&+& \frac{\alpha_s}{2\pi}\bigg [ \log \left (\frac{\mu^2}{M^2} \right )-\frac{1}{\hat{\epsilon}} \bigg ]\int_{z_1}^{1-z_2} \frac{du}{u(1-u)}\Bigg [\hat{P}_{ji}^k(u)\, D_j^{H_1}\Big(\frac{x}{u}\Big ) D_k^{H_2}\Big (\frac{y}{1-u}\Big )\Bigg ].
\end{eqnarray}

 The factorization relation in Eq.(\ref{ecuaciones de renormalizacion DD}) contains two terms with different origins. The first one just relates the factorized and bare double-fragmentation functions, and corresponds to the standard factorization procedure for the emission of collinear partons in the double fragmentation part of the cross section, exactly as it occurs for one-hadron production. The second `inhomogeneous' term relates the single and double-fragmentation  functions and is  needed to absorb the remaining singularities discussed above.
 
 Rewriting the bare distributions in terms of the factorized ones in Eq.(\ref{sigmaDxD general}), and fixing $M^2=Q^2$,  we obtain the final NLO  expression for the factorized  cross section for the production of two hadrons as
\begin{eqnarray}\label{sigmaH1H2 qqg resultado renormalizado}
{\frac{d\sigma^{H_1H_2}}{dz_1dz_2}}^{(M)}  &=&  3 \sigma_0\int_{z_1}^{1-z_2}\frac{dx_1}{x_1} \int_{1-x_1}^1  \frac{dx_2}{x_2} \sum_{q}e_q^2 \Bigg \{ \frac{\alpha_s}{2\pi} f_{qq A}^{(M)}(x_1,x_2) \nonumber \\ \nonumber \\ & &  \bigg [ D_q^{H_1 (NLO)} \Big (\frac{z_1}{x_1}, Q^2 \Big ) D_{\bar q}^{H_2(NLO)} \Big (\frac{z_2}{x_2},Q^2 \Big ) +  D_q^{H_2 (NLO)} \Big (\frac{z_2}{x_2},Q^2 \Big ) D_{\bar q}^{H_1(NLO)} \Big (\frac{z_1}{x_1},Q^2 \Big ) \bigg ]  \nonumber \\ \nonumber\\ &  +   &\frac{\alpha_s}{2\pi} f_{qg A}^{(M)}(x_1,x_2)  \bigg [ D_q^{H_1(NLO)} \Big (\frac{z_1}{x_1},Q^2 \Big )  +  D_{\bar q}^{H_1 (NLO)} \Big (\frac{z_1}{x_1},Q^2 \Big )  \bigg ]D_{g}^{H_2 (NLO)} \Big (\frac{z_2}{x_2},Q^2 \Big )\nonumber \\ \nonumber \\ &  +  & \frac{\alpha_s}{2\pi} f_{gq A}^{(M)}(x_1,x_2) \bigg [ D_q^{H_2 (NLO)} \Big (\frac{z_2}{x_2},Q^2 \Big )  +  D_{\bar q}^{H_2(NLO)} \Big (\frac{z_2}{x_2},Q^2 \Big )  \bigg ]D_g^{H_1(NLO)} \Big (\frac{z_1}{x_1},Q^2 \Big )\Bigg \} +  \nonumber \\ \nonumber \\
 &  +    & 3 \sigma_0 \int_{1-z_2}^1 \frac{dx_1}{x_1} \int_{z_2}^1 \frac{dx_2}{x_2} \sum_{q}e_q^2 \Bigg \{ \Big [\delta(1-x_1)\delta(1-x_2)+\frac{\alpha_s}{2\pi}  f_{qq B}^{(M)}(x_1,x_2) \Big ]\nonumber \\ \nonumber \\ &      & \bigg [ D_q^{H_1(NLO)} \Big (\frac{z_1}{x_1},Q^2 \Big ) D_{\bar q}^{H_2(NLO)} \Big (\frac{z_2}{x_2},Q^2 \Big ) +  D_q^{H_2 (NLO)} \Big (\frac{z_2}{x_2},Q^2 \Big ) D_{\bar q}^{H_1(NLO)} \Big (\frac{z_1}{x_1},Q^2 \Big ) \bigg ]  \nonumber \\ \nonumber\\ &  +  &\frac{\alpha_s}{2\pi}  f_{qg B}^{(M)}(x_1,x_2) \bigg [ D_q^{H_1(NLO)} \Big (\frac{z_1}{x_1},Q^2 \Big )  +  D_{\bar q}^{H_1 (NLO)} \Big (\frac{z_1}{x_1},Q \Big )  \bigg ]D_{g}^{H_2 (NLO)} \Big (\frac{z_2}{x_2},Q^2 \Big )  \nonumber \\ \nonumber \\ &   +   & \frac{\alpha_s}{2\pi}  f_{gq B}^{(M)}(x_1,x_2) \bigg [ D_q^{H_2 (NLO)} \Big (\frac{z_2}{x_2},Q^2 \Big )  +  D_{\bar q}^{H_2 (NLO)} \Big (\frac{z_2}{x_2},Q^2 \Big )  \bigg ]D_g^{H_1 (NLO)} \Big (\frac{z_1}{x_1},Q^2 \Big )\Bigg \}\nonumber  \\ \nonumber \\  &   +  & 3 \sigma_0 \int_{z_1+z_2}^1 \frac{dx}{x^2}\sum_{q}e_q^2 \Bigg \{ \Big[\delta(1-x)+ \frac{\alpha_s}{2\pi} f^{(M)}_q(x) \Big ] \bigg [ DD_q^{H_1 H_2 (NLO)}\Big (\frac{z_1}{x},\frac{z_2}{x},Q^2 \Big )\nonumber \\ \nonumber \\ & +  & DD_{\bar q}^{H_1 H_2 (NLO)}\Big (\frac{z_1}{x},\frac{z_2}{x},Q^2 \Big )\bigg ]+ 2 \frac{\alpha_s}{2\pi} f^{(M)}_g(x) DD_g^{H_1 H_2 (NLO)}\Big (\frac{z_1}{x},\frac{z_2}{x},Q^2 \Big )\Bigg \}.
\end{eqnarray}

The longitudinal contribution is obtained by replacing $f^{(M)} \rightarrow f^{(L)}$ and omitting the corresponding LO  terms proportional to $\delta(1-x_1)\delta(1-x_2)$ and $\delta(1-x)$.
 
While the dependence on the momentum fractions of the double-fragmentation functions can not be computed within perturbative QCD, the factorization scale dependence is driven by the evolution equations.
 As for the ordinary  fragmentation functions, these equations 
can be  obtained by requiring that the bare functions $DD_i^{H_1H_2}$ do not depend on the factorization scale 
 \begin{equation}
\frac{d}{d \log M^2}DD_i^{H_1H_2}(x,y)=0.
\end{equation}

Replacing  $DD_i^{H_1H_2}(x,y)$ from  Eq.(\ref{ecuaciones de renormalizacion DD}) we obtain:

\begin{eqnarray}\label{evolucion para DD}
\frac{d}{d \log M^2}DD_i^{H_1H_2(NLO)}(x,y,M^2) &=&\frac{\alpha_s}{2\pi}\int_{x+y}^1 \frac{du}{u^2} P_{ji}(u) DD_j^{H_1H_2(NLO)}\Big (\frac{x}{u},\frac{y}{u}, M^2 \Big ) \\ \nonumber \\ & + &  \frac{\alpha_s}{2\pi}\int_{z_1}^{1-z_2} \frac{du}{u(1-u)}\Bigg [\hat{P}_{ji}^k(u)\, D_j^{H_1}\Big(\frac{x}{u}\Big ) D_k^{H_2}\Big (\frac{y}{1-u}\Big )\Bigg ].\nonumber
\end{eqnarray}

\begin{figure}[htb]
\begin{center}
\begin{tabular}{c}
\epsfxsize=13.0truecm
\epsffile{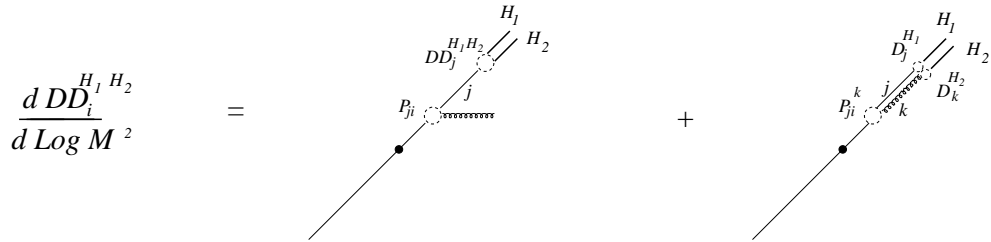}\\
\end{tabular}
\end{center}
\caption{\label{Evol}
{\em Graphic representation for the evolution of the $DD_i^{H_1H_2}$. The first diagram corresponds to the first term in Eq.(\ref{evolucion para DD}). The last one to the inhomogeneous term. }}
\end{figure}

  The first term  in the right hand-side  corresponds to the usual homogeneous evolution of the fragmentation functions $D_i^H$. It indicates that the probability of obtaining the  hadrons  $H_1$ and $H_2$ from the parton $i$ is affected by the possibility of the emission of a parton  $j$ with momentum fraction $u$, which can produce two hadrons by  a double fragmentation.  The second term, on the other hand, is inhomogeneous and it is not  present in the evolution equations of the $D_i^H$ functions. It corresponds to the case of a the parton $i$  that evolves emitting the partons $j$ and $k$ with fractions $u$ and $1-u$ respectively, which can also give origin to two hadrons, but now by means of the mechanism of single fragmentation form each one of them. Both terms are represented symbolically in the Fig.\ref{Evol}. These equations fully agree with the ones originally  proposed in \cite{mencion de las DD, evolucion de las DD}.

The presence of these two terms in the evolution equations evidences the fact that, within the precision of any possible detector, it is physically impossible to determine which  mechanism, either single or double  fragmentation, has been responsible for the production of two hadrons when they are found in the same  jet.  In the same way, and beyond LO accuracy, only the sum of the two terms in 
Eq.(\ref{ee H1H2 convolucion}) associated to each one of the mechanisms that contributes to the cross section, has a physical meaning. In this sense, the similarity with the situation of the {\em fracture functions}\cite{fractura:c} in DIS appears again.

\section{Conclusion.}

In this work, the cross section for the production of  two hadrons in $e^+e^-$ annihilation  is calculated to order $\alpha_s$ considering events that include the possibility that both  hadrons appear in the same jet. For this purpose it is necessary to extend the fragmentation model  including a new type of functions, the  double-fragmentation functions $DD_i^{H_1H_2}$, that  describe the transition of a parton into two hadrons. These functions, along with the single-fragmentation function $D_i^H$, allow an  unified treatment  for the description of two-hadron production in $e^+e^-$ annihilation.
  
 While  at leading order the $DD_i^{H_1H_2}$  functions  are necessary to contemplate the possibility of the double fragmentation,  at next-to-leading order and beyond,  they are required to perform the factorization of divergences that  cannot be absorbed in the  single-fragmentation functions. As a result, they 
obey the inhomogeneous evolution equations  in Eq.(\ref{evolucion para DD}),  where the two mentioned mechanisms of fragmentation  are involved.  
     We showed,  for the first time, that introducing the double-fragmentation functions the usual factorization procedure can be enlarged consistently for the production of two hadrons to order $\alpha_s $, reobtaining  the  evolution equations originally proposed in \cite{mencion de las DD, evolucion de las DD}. 

\section*{Acknowledgements}

We would like to thank Alejandro Daleo and Rodolfo Sassot for many valuable discussions and comments.

\appendix

\section{Appendix}
\begin{appendix}
The NLO($\overline{MS}$) corrections to the single fragmentation term are given by
\begin{eqnarray}
 f_{q \bar q A}^{(M)}(x_1,x_2) & = &  \frac{4}{3}\frac{x_1^2+x_2^2}{(1-x_1)(1-x_2)_{+[0, \,\underline{1}]}}  + \Bigg [ \hat{P}_{qq}(x_1) \log \Big [ (1-x_1)x_1 \Big ]+ \frac{4}{3} (1-x_1) \Bigg ]\delta(1-x_2),
\nonumber \\ \nonumber \\
 f_{q \bar q B}^{(M)}(x_1,x_2) & = &  \frac{4}{3}\frac{x_1^2+x_2^2}{(1-x_1)_{+[0, \,\underline{1}]}\,(1-x_2)_{+[0, \,\underline{1}]}} +\frac{4}{3} \delta(1-x_1)\delta(1-x_2)(\pi^2-8) +
\nonumber \\ \nonumber \\ 
& + &  \Bigg[ \frac{4}{3} (1+x_1^2)\bigg (\frac{\log(1-x_1)}{1-x_1} \bigg )_{+[0, \,\underline{1}]} + \frac{4}{3}(1-x_1)+ \hat{P}_{qq}(x_1)\log x_1 \Bigg ]\delta(1-x_2)
\nonumber \\ \nonumber \\ 
& + &  \Bigg[ \frac{4}{3} (1+x_2^2)\bigg (\frac{\log(1-x_2)}{1-x_2} \bigg )_{+[0, \,\underline{1}]} + \frac{4}{3}(1-x_2)+ \hat{P}_{qq}(x_2)\log x_2 \Bigg ]\delta(1-x_1), \nonumber \\ \nonumber \\
f_{qg A}^{(M)}(x_1,x_2) & = &  \frac{4}{3}\frac{x_1^2 + (2-x_1+ x_2)^2 }{(1-x_1)(x_1+x_2-1)_{+[\underline{1-x_2}, \,1]}  }+ \Bigg [\hat{P}_{qq}(x_1) \Big [\log(1-x_1)x_1^2 \Big ]\nonumber \\ \nonumber \\ & + &  \frac{4}{3} (1-x_1)  \Bigg ]\delta(x_1+x_2-1),
\nonumber \\ \nonumber \\
 f_{qg B}^{(M)}(x_1,x_2)  & = & \frac{4}{3}\frac{x_1^2 + (2-x_1-x_2)^2}{(1-x_1)_{+[0, \,\underline{1}]}\,(x_1 + x_2-1)} + \Bigg [\hat{P}_{gq}(x_2)\Big [\log(1-x_2)x_2 \Big ] +  \frac{4}{3}x_2 \Bigg ]\delta(1-x_1),\nonumber \\  \nonumber \\
 f_{gq A}^{(M)}(x_1,x_2)  & = &  \frac{4}{3}\frac{x_2^2 + (2-x_1-x_2)^2}{(1-x_2)_{+[0, \,\underline{1}]}\,(x_1+x_2-1)_{+[\underline{1-x_2}, \,1]}} + \Bigg[\hat{P}_{gq}(x_1) \Big [\log(1-x_1)x_1 \Big ] + \frac{4}{3}x_1\Bigg ]\delta(1-x_2)\nonumber \\ \nonumber \\ & + & \Bigg [\hat{P}_{gq}(x_1)\Big [\log (1-x_1)^2x_1 \Big ] + \frac{4}{3}x_1   \Bigg ]  \delta(x_1+x_2-1), 
\nonumber \\ \nonumber \\  
 f_{gq B}^{(M)}(x_1,x_2) & = &  \frac{4}{3}\frac{x_2^2 + (2-x_1-x_2)^2}{(1-x_2)_{+[0, \,\underline{1}]}\,(x_1 + x_2-1)} + \Bigg [\hat{P}_{gq}(x_1)\Big [\log(1-x_1)x_1 \Big ] + \frac{4}{3}x_1 \Bigg ]\delta(1-x_2), \nonumber \\ \nonumber \\ 
{f_{q \bar q}}_{A/B}^{(L)}(x_1,x_2) & = & \frac{8}{3}   \Big (\frac{x_1+x_2-1}{x_2^2} \Big ),\nonumber \\ \nonumber \\  
{f_{qg}}_{A/B}^{(L)}(x_1,x_2) & = &  \frac{16}{3}   \Big (\frac{1-x_2}{x_2^2} \Big ),\nonumber \\ \nonumber \\  
{f_{gq}}_{A/B}^{(L)}(x_1,x_2) & = &   \frac{8}{3}   \Big (\frac{1-x_1}{x_2^2} \Big ).\nonumber 
\end{eqnarray}
The corresponding corrections to double fragmentation are
\begin{eqnarray}
f^{(M)}_q(x) & = & \frac{4}{3}  \Bigg [(1+x^2)\left (\frac{\log1-x}{1-x}\right)_{+[0, \,\underline{1}]} + 2 \left ( \frac {1+x^2}{1-x}\right )\log x - \frac{3}{2}\frac{1}{(1-x)_{+[0, \,\underline{1}]}+}   -  \frac{3}{2}x + \frac{5}{2}     \nonumber  \\ \nonumber \\ &+& \left (\frac{2\pi^2}{3} - \frac{9}{2}\right )\delta (1-x) \Bigg ],
\nonumber \\ \nonumber \\
 f^{(M)}_g(x)  & = & \hat{P}_{gq}(x) \Big [ \log (1-x) x^2  \Big ], \label{fg}
\nonumber\\ \nonumber \\
f_q^{(L)}(x) & = &  \frac{4}{3},\nonumber \\ \nonumber\\ 
f_g^{(L)}(x) & = &   \frac{8}{3}   \Big (\frac{1-x}{x} \Big ). \nonumber
\end{eqnarray}

\end{appendix}


\begin{thebibliography}{99}
\bibitem{altarelli B160} G. Altarelli, R.K. Ellis, G. Martinelli, So-Young Pi.  Nucl. Phys. B 160. (1979), p. 301.
\bibitem{Rijken Neerven} P.J. Rijken, W.L. van Neerven.   Nucl. Phys. B 487 (1997),  p. 233. 


\bibitem{mencion de las DDa} K. Konishi, A. Ukawa, G. Veneziano. Phys. Lett. B 78  (1978), p.243.
\bibitem{mencion de las DD} K. Konishi, A. Ukawa, G. Veneziano.  Nucl. Phys. B 157 (1979), p. 45.
\bibitem{vendramin 66} I. Vendramin, Il Nuovo Cimento   A 66    (1981), p. 339.

\bibitem{evolucion de las DD} U.P. Sukhatme, K.E. Lassila,  Phys. Rev. D 22 (1980), p. 1184


\bibitem{fractura:c} L. Trentadue, G. Veneziano. Phys. Lett. B 323 (1994), p. 201. 
\bibitem{fractura} D. Graudenz.  Nucl. Phys. B 432 (1994), p. 351.
\bibitem{fractura:b} D. de Florian, C.A. Garc\'\i a Canal, R. Sassot. Nucl. Phys. B 470 (1996), p. 195.

\bibitem{fractura:d} G. Camici, M. Grazzini, L. Trentadue.  Phys. Lett. B 439 (1998), p. 382. 

\bibitem{Grazzini:1997ih} M.~Grazzini, L.~Trentadue and G.~Veneziano,
Nucl.\ Phys.\ B 519 (1998), p. 394.

\bibitem{regularizacion} C. G. Bollini, J.J. Giambiagi. Nuovo Cimento B 12  (1972), p. 20.
\bibitem{regularizacion:a} G.'t Hooft, M. Veltman. Nucl. Phys. B 44  (1972), p. 189.


\bibitem{altarelli B126} G. Altarelli, G. Parisi. Nucl. Phys. B 126 (1977),  p. 298. 

\end{thebibliography}
\end{document}